\documentclass[12pt]{article}
\usepackage{amsmath}

\usepackage{amssymb}

\numberwithin{equation}{section}

\begin{document}

\begin{center}
 La Probabilidad de la Mec\'anica Cu\'antica: \\
 Una Introducci\'on en noventa minutos\\
 Stephen Bruce Sontz\\
Centro de Investigaci\'on en Matem\'aticas, A.C.\\
(CIMAT)\\
Guanajuato, Gto., M\'exico\\
email: sontz@cimat.mx

\vskip 0.4cm \noindent
    \textbf{Resumen}

\end{center}

\vskip 0.1cm \noindent

 Einstein dec\'ia en referencia a la mec\'anica cu\'antica:
``The old one does not play dice'' - y es cierto, porque la
 probabilidad que se usa en la mec\'anica cu\'antica no es la probabilidad
 cl\'asica de juegos como los dados. Es el primer ejemplo
hist\'oricamente hablando de
una probabilidad que se llama no cl\'asica o no conmutativa.
Vamos a presentar lo b\'asico de esta probabilidad cu\'antica a un nivel introductorio para
estudiantes de licenciatura con conocimiento de \'algebra lineal.
Usamos la teor\'ia de probabilidad con un espacio finito
como analog\'ia para el caso de dimensi\'on finita de la
probabilidad cu\'antica.
Adem\'as hay un ap\'endice breve sobre  un tema relacionado (qubits)
para indicar que las ideas presentadas aqu\'i tienen otras aplicaciones.
Conocimiento de la mec\'anica cu\'antica no es necesario.

\section{Probabilidad Cl\'asica - Caso Finito}

Antes de empezar notamos que
no demostramos todo porque el art\'iculo est\'a basado en una conferencia
de noventa minutos.
Entonces, cada afirmaci\'on 
sin demostraci\'on o con demostraci\'on incompleta es un ejercicio para el lector.

\vskip 0.1cm \noindent
Un caso especial de
la Probabilidad Cl\'asica consta de:

\begin{enumerate}
 \item
 un conjunto $\Omega$ que es \textit{finito} y \textit{no vac\'io},
\item
\textit{todos} los subconjuntos $E \subset \Omega$
 que
se llaman \textit{Eventos},
\item una \textit{Funci\'on de Probabilidad}
$
         P :  E \mapsto  P(E) \in [0,1]
$
es decir, \\
$0 \le P(E) \le 1$ para cada evento $E \subset \Omega$.
\end{enumerate}
Se dice que $P(E)$ es la \textit{Probabilidad del Evento} $E$, si
 las propiedades siguientes se cumplen:
\vskip 0.1cm \noindent \hspace{0.8cm}
 1. $P(\emptyset)=0$ \hspace{0.2cm} (donde $\emptyset$ es el conjunto vac\'io),
 \hspace{0.9cm}
2. $P(\Omega)=1$,
\vskip 0.2cm \noindent \hspace{0.8cm}
3. $P(E_1 \cup E_2 \cup \cdots \cup E_k ) = P(E_1) + P(E_2) + \cdots + P(E_k) $
donde \\
\hspace*{1.3cm}
$E_1, \dots , E_k$ son \textit{Eventos Disjuntos}, o sea,
$E_i \cap E_j = \emptyset$ si $i \ne j$.
\vskip 0.2cm \noindent
As\'i es la teor\'ia (Fermat, Pascal; 1600's) desarrollada
para estudiar juegos como las cartas.
Hoy d\'ia tiene muchas aplicaciones en ciencias como biolog\'ia,
ingenier\'ia, f\'isica, y por cierto matem\'aticas entre otras.

No vamos a ver en este art\'iculo la teor\'ia de \textit{Probabilidad Cl\'asica} debida a Kolmogorov
(1933, v\'ease \cite{KOLM}) para el caso cuando $\Omega$ es \textit{infinito},
dado que tiene un nivel m\'as avanzado con muchos detalles t\'ecnicos.

Un comentario clave
aqu\'i es que hay experimentos y observaciones con la propiedad de
que cuando est\'an hechos bajo condiciones iniciales iguales (o esencialmente iguales)
no siempre dan el mismo resultado.
Es decir, hay un conjunto de dos o m\'as resultados que son las mediciones posibles.
Son los casos para los cuales hay necesidad de una teor\'ia
que no sea \textit{determinista} (que describe una teor\'ia que predice exactamente un solo resultado
para cada experimento u observaci\'on).
En cambio una teor\'ia que describe situaciones donde hay m\'as de un resultado posible
se llama una \textit{Teor\'ia de Probabilidad} que sea la teor\'ia cl\'asica de
Kolmogorov u otra.

Entonces son importantes
las estructuras matem\'aticas en una teor\'ia de probabilidad
que corresponden a los n\'umeros reales \textit{medidos}
(u \textit{observados}) en experimentos.
En la probabilidad cl\'asica, esta estructura se llama
una \textit{Variable Aleatoria}  y es por definici\'on una funci\'on con valores reales:
$$
         X : \Omega \to \mathbb{R}.
$$

La idea detr\'as de esta definici\'on es que los n\'umeros reales en el rango (o imagen) de $X$,
$\, \mathrm{Ran}(X):=\{ \lambda \in \mathbb{R}  \, | \, \exists \, \omega \in \Omega, \, \lambda = X(\omega) \}$,
son todos los valores medibles posibles de la cantidad experimental correspondiente a $X$.

Se define el \textit{Valor Esperado} (o m\'as bien el \textit{Valor Promedio})
de $X$ con respecto a $P$ por
$$
\left\langle X \right\rangle  := \sum_{\omega \in \Omega} P( \omega ) X(\omega)
\quad \mathrm{donde} \quad P(\omega) \equiv P(\{ \omega \}).
$$
Notamos que
el conjunto $
 \{ \, X ~ \big\vert  ~ X : \Omega \to \mathbb{R} \, \}$
 de todas las variables aleatorias
es un espacio vectorial sobre $\mathbb{R}$ de dimensi\'on finita $n = \mathrm{card}(\Omega) \ge 1$.
Tambi\'en tiene un producto conmutativo dado por la multiplicaci\'on usual
de funciones:
$$
        X Y (\omega) = X \cdot Y (\omega) := X(\omega) Y(\omega)
$$
para cada $\omega \in \Omega $ donde $X,Y$ son variables aleatorias.

\section{Mec\'anica Cu\'antica en Dimensi\'on Finita}

\noindent
La \textit{Mec\'anica Cu\'antica} de Pauli y Wigner
($\sim$1930) empieza con el espacio vectorial
$$
         \mathbb{C}^{n} = \{ z = (z_1, z_2, \dots , z_n ) ~|~ z_j \in \mathbb{C}
            \, \, \, \mathrm{para}\, 1 \le j \le n \}
$$
de \textit{dimensi\'on finita}  $n \ge 1$  sobre los n\'umeros complejos $\mathbb{C}$ con su
\textit{producto interior}
$$
 \left\langle z,w \right\rangle := \sum_{j=1}^n z_j^* w_j.
$$
Aqu\'i $z = (z_1, \dots , z_n) \in \mathbb{C}^n,\, w = (w_1, \dots , w_n) \in \mathbb{C}^n, \,
\beta^*$ es la \textit{conjugada compleja} de $\beta \in \mathbb{C}$.
Tambi\'en, $\mathbb{C}^n$ tiene una \textit{norma} $|| \cdot ||$  dada por
$$
   || z ||^2 = \left\langle z, z \right\rangle = \sum_{j=1}^n z_j^* z_j =
\sum_{j=1}^n |z_j|^2 \quad
   \mathrm{para~cada~} z \in \mathbb{C}^n.
$$

No vamos a ver en este art\'iculo
la teor\'ia de la \textit{Mec\'anica Cu\'antica} de Heisenberg y
Schr\"odinger (1925-26) para el caso de
dimensi\'on \textit{infinita}, dado que tiene un nivel m\'as avanzado
con muchos detalles t\'ecnicos.
Al igual que la probabilidad cl\'asica general de Kolmogorov,
es importante pero no tenemos que estudiarla por lo pronto para nuestras metas.
Por cierto, el estudiante interesado tiene que aprender tarde o temprano  en el caso general
la probabilidad cl\'asica, la mec\'anica cu\'antica y la probabilidad cu\'antica
(entre otros temas).
Por lo tanto recomendamos fuertemente la \textit{lectura}.

Para definir las estructuras matem\'aticas en la mec\'anica cu\'antica que corresponden a
cantidades medidas en experimento, introducimos
el conjunto de \textit{matrices} complejas $n \times n $ (con $n$ entero, $n \ge 1$)
$$
    \mathrm{MAT} (n;\mathbb{C} ) := \{ A = (A_{jk}) \mathrm{,~matriz~} n \times n, \,
                                      A_{jk} \in \mathbb{C}, \, 1 \le j,k \le n  \}.
$$
\textbf{N.B.} Cada matriz $A$ define un mapeo lineal
$A : \mathbb{C}^{n} \to \mathbb{C}^{n} $ en esta manera:
$(Az)_j = \sum_{k=1}^n A_{jk} z_k $.
(Usamos un abuso com\'un de notaci\'on; el s\'imbolo $A$ denota a la matriz
y al mapeo lineal correspondiente.)
Rec\'iprocamente cada mapeo lineal $ \mathbb{C}^{n} \to \mathbb{C}^{n} $
viene en la manera indicada de una matriz \'unica en $ \mathrm{MAT} (n;\mathbb{C} )$.
Resulta que $\, \mathrm{MAT} (n;\mathbb{C} ) $
es un espacio vectorial
de dimensi\'on $n^2$ sobre $\mathbb{C}$.
Cuenta con el producto usual de matrices que es \textit{no~conmutativo}
si $n \ge 2$.
Por ser un espacio vectorial con producto compatible, resulta que
$\mathrm{MAT} (n;\mathbb{C} )$ es un \textit{\'algebra} sobre $\mathbb{C}$.

Toda matriz $A$ tiene una \textit{matriz adjunta} $A^*$ donde
$ (A^*)_{jk} := (A_{kj})^* $ para $ 1 \le j,k \le n$.
Es la matriz transpuesta conjugada.
Resulta que $A^*$ es la matriz \'unica tal que
$\left\langle A^*z, w \right\rangle = \left\langle z, A w \right\rangle $
para todos vectores $z, w \in \mathbb{C}^n$.

Si $A = A^*$ se dice que $A$ es \textit{Auto-Adjunta}
(o \textit{Hermitiana}).
Notaci\'on:
$$
\mathrm{HERM}(n) := \{ A \in \mathrm{MAT}(n; \mathbb{C}) ~|~ A = A^* \}.
$$
 Las matrices auto-adjuntas corresponden a muchas de
las cantidades medidas en experimentos
con sistemas cu\'anticos.
Vamos a ver m\'as adelante con todo detalle la correspondencia, pero por
lo pronto cabe subrayar que es por eso que las matrices
auto-adjuntas son importantes en f\'isica cu\'antica.

\vskip .1cm \noindent
\textbf{Ejercicio:}
 $\mathrm{HERM}(n) $ es un espacio vectorial sobre los n\'umeros reales $\mathbb{R}$
y \textit{no} lo es sobre los n\'umeros complejos $\mathbb{C}$.
No es una sub\'algebra de $\mathrm{MAT}(n;\mathbb{C})$ si $n \ge 2$.
Adem\'as: $\, \dim_{\mathbb{R} } \mathrm{HERM}(n) = n^2 $. $ \quad \blacksquare$

\vskip .1cm \noindent
\textbf{Ejemplo:}
$\mathrm{HERM}(2) $ tiene dimensi\'on $2^2 =4$ y una base est\'a dada por
las tres \textit{Matrices de Pauli}
\begin{equation*}
 \sigma_1 = \left( \begin{array}{cr} 0 & 1 \\ 1 & 0 \end{array} \right),
\quad
\sigma_2 = \left( \begin{array}{cr} 0 & -\mathrm{i} \\ \mathrm{i} & 0 \end{array} \right),
\quad
\sigma_3 = \left( \begin{array}{cr} 1 & 0 \\ 0 & -1 \end{array} \right)
\end{equation*}
 y la identidad, $\, I = \left( \begin{array}{cc} 1 & 0 \\ 0 & 1 \end{array} \right)$.
(Aqu\'i $\mathrm{i} = \sqrt{-1} $.)
Unas propiedades de ellas:
$$
 \sigma_1 \sigma_2 = \mathrm{i} \sigma_3 = - \sigma_2 \sigma_1, \qquad
\sigma_2 \sigma_3 = \mathrm{i} \sigma_1 = - \sigma_3 \sigma_2, \qquad
\sigma_3 \sigma_1 = \mathrm{i} \sigma_2 = - \sigma_1 \sigma_3.
$$

\section{Teor\'ia Espectral}

Vamos a usar un teorema fundamental del \'algebra lineal:
cada matriz auto-adjunta tiene una \textit{diagonalizaci\'on}.
La siguiente es una manera para enunciar este teorema.
Una referencia para esta secci\'on es el texto \cite{HALM} por Halmos.

\vskip .1cm \noindent
Sea $A$ una matriz en $\mathrm{MAT}(n; \mathbb{C}) $.
Se define su \textit{Polinomio Caracter\'istico} por
$$
p_{A}(\lambda) := \det (\lambda I -A).
$$
Es un polinomio de grado $n \ge 1$ en $\lambda$ con coeficientes complejos,
donde $\lambda$ es una variable compleja.
 (Aqu\'i $I \in \mathrm{MAT}(n; \mathbb{C})$ es la matriz identidad y det es el determinante.)
Podemos escribir el conjunto de las ra\'ices complejas \textit{distintas} de $p_{A}(\lambda)$ como
$$
\mathrm{SPEC} (A) := \{ \lambda_1, \lambda_2, \dots , \lambda_k \} \subset \mathbb{C}
$$
con $1 \le k \le n$
(por el Teorema Fundamental de \'Algebra).
Decimos que $\mathrm{SPEC} (A) $ es el \textit{Espectro} de $A$
 y que
$\lambda_1, \lambda_2, \dots , \lambda_k  $ son los \textit{Eigenvalores} de $A$.
Hay varias caracterizaciones de un eigenvalor.
Las siguientes afirmaciones son equivalentes:
\begin{itemize}
 \item  $\beta \in \mathbb{C}$ es un eigenvalor de $A$.
\item  $p_{A}(\beta)=0$.
\item $\det (\beta I -A) =0 $.
\item La matriz $\beta I - A $ no es invertible.
\item El \textit{subespacio} $\mathrm{ker} (\beta I - A) $ no es cero.\\
(Aqu\'i $\mathrm{ker}(B) :=  \{ z \in \mathbb{C}^n  \, | \, Bz=0 \}$
para $B: \mathbb{C}^n \to \mathbb{C}^n$ lineal.)
\item Existe $ z \in \mathbb{C}^n$ con $z \ne 0 \,$ y $\, A z = \beta z $.\\
(En tal caso $z$ se llama un \textit{eigenvector de $A$ asociado a} $\beta$.)
\end{itemize}
Quiz\'as la \'ultima afirmaci\'on sea la propiedad m\'as familiar para el lector.
Sin embargo, usaremos la pen\'ultima.
Expl\'icitamente, definimos los subespacios
$$V_j := \ker (\lambda_j I - A) \ne 0 \quad \mathrm{para} \quad 1 \le j \le k.
$$
Entonces:
$z \in V_j \Longleftrightarrow  (\lambda_j I - A)  z =0 \Longleftrightarrow  A z = \lambda_j z$,
o sea, la acci\'on de $A$ en el subespacio $V_j$ es igual a la acci\'on de multiplicaci\'on por $\lambda_j$.
Se dice que $A$ \textit{tiene el valor} $\lambda_j$ en el subespacio $V_j$.

Si adem\'as $A$ es auto-adjunta, entonces tenemos:
\begin{itemize}
 \item Todos sus eigenvalores son reales.
\item Los subespacios $V_j$ son ortogonales,
o sea,\\
 $\left\langle  z_i, z_j \right\rangle =0$ si $z_i \in V_i $, $z_j \in V_j $ para $i \ne j$.
\item $\mathbb{C}^n = V_1 \oplus V_2 \oplus \cdots \oplus V_k. \quad $ (\textit{Suma Directa})
\end{itemize}
Dejamos la tercera afirmaci\'on como ejercicio.
Para probar las dos primeras afirmaciones, tomamos $z_i \in V_i, z_j \in V_j$.
Entonces para $ 1 \le i,j \le k$ tenemos
$$
\lambda_j \left\langle  z_i, z_j \right\rangle =  \left\langle  z_i, \lambda_j z_j \right\rangle =
 \left\langle z_i, A z_j \right\rangle =
\left\langle A z_i, z_j \right\rangle =  \left\langle \lambda_i z_i, z_j \right\rangle   =
\lambda_i^* \left\langle  z_i, z_j \right\rangle.
$$
Tomando $i =j$ y $z_i \ne 0 $ y usando $\left\langle  z_i, z_i \right\rangle \ne 0$,
concluimos que $\lambda_i = \lambda_i^* $, es decir $\lambda_i$ es real.
Luego tomando $ i \ne j$, tenemos que $(\lambda_i - \lambda_j) \left\langle  z_i, z_j \right\rangle =0 $,
que implica $ \left\langle  z_i, z_j \right\rangle =0 $ porque
$\lambda_i \ne \lambda_j$.

\noindent
Entonces, al formar la uni\'on
 de una base ortonormal de $V_1$,
 de una base ortonormal de $V_2$,
 $ \dots$,
 de una base ortonormal de $V_k$,
obtenemos una base ortonormal de $\mathbb{C}^n$ con la propiedad
que el mapeo lineal
$$
A = A^* : \mathbb{C}^n \to \mathbb{C}^n
$$
tiene una matriz diagonal en esta base ortonormal nueva.
Esto se llama la \textit{diagonalizaci\'on} de la matriz auto-adjunta $A$.
Se encuentran a lo largo de la diagonal
$\dim V_1 \ge 1 $ ocurrencias de $\lambda_1$,
$\dim V_2 \ge 1 $ ocurrencias de $\lambda_2, \dots,$
$\dim V_k \ge 1 $ ocurrencias de $\lambda_k$.

Vamos a seguir un paso m\'as adelante.
En lugar de usar subespacios de $\mathbb{C}^n$
vamos a usar proyectores ortogonales.
Sea $W \subset \mathbb{C}^n$ un subespacio de $\mathbb{C}^n$.
Resulta que podemos \textit{descomponer} $\mathbb{C}^n$ como una \textit{suma directa},
$$
  \mathbb{C}^n = W \oplus W^\perp,
$$
donde el subespacio
$$
W^\perp : = \{  v \in \mathbb{C}^n ~|~ \left\langle v , w \right\rangle =0
\mathrm{~para~todo~} w \in W \}
$$
se llama el \textit{complemento ortogonal} de $W$
en $\mathbb{C}^n$.

Entonces, al subespacio $W$ asociamos el mapeo lineal $E_W : \mathbb{C}^n \to \mathbb{C}^n$
definido por
$$
   E_W z := w
$$
para $z \in \mathbb{C}^n$
con $z = w + v \,$ su descomposici\'on \'unica con $w \in W$, $v \in W^\perp$.
Resulta que
$$
          E_W = E_W^* = E_W^2 \qquad \mathrm{y} \qquad \mathrm{Ran}(E_W) = W.
$$
(Aqu\'i el rango de $B : \mathbb{C}^n \to \mathbb{C}^n$ es
$\mathrm{Ran}(B): = \{ \tilde{z} \in \mathbb{C}^n \, | \, \exists z \in \mathbb{C}^n, \tilde{z} = Bz \}$.)
Si $E : \mathbb{C}^n \to \mathbb{C}^n$ es un mapeo lineal que satisface $E = E^* = E^2$, se llama un
\textit{Proyector Ortogonal}.
Y se dice que $E_W$ es el \textit{Proyector Ortogonal sobre} $W$.

Rec\'iprocamente, para cada proyector
ortogonal $E$ tenemos que
$E = E_W $ para un subespacio \'unico, a saber,
 $W = \mathrm{Ran}(E) $.
En fin de cuentas, hemos obtenido una correspondencia uno-a-uno y sobre (es decir, una \textit{biyecci\'on})
entre el conjunto de todos los subespacios $W$ de $\mathbb{C}^n$
y el conjunto de todos los proyectores ortogonales en $\mathrm{MAT}(n; \mathbb{C})$:
$\quad W \longleftrightarrow E_W$.

Por fin, podemos enunciar el teorema de diagonalizaci\'on.
Este teorema usa el concepto de proyector ortonormal para expresar
los resultados que ya hemos visto anteriormente.

\vskip .1cm \noindent
\textbf{Teorema:}
Si $A \in \mathrm{MAT}(n,\mathbb{C})$ es una matriz auto-adjunta,
entonces existe un entero $k$ con $1 \le k \le n$ y
existen proyectores ortogonales $E_1 \ne 0$,
$ \dots $, $ E_k \ne 0$
en $\mathrm{MAT}(n,\mathbb{C}) $ y
n\'umeros reales \textit{distintos}  $\lambda_1,  \dots , \lambda_k$ tales que
\begin{gather*}
 E_i E_j = 0 \quad \mathrm{si} \quad i \ne j, \\
I = E_1 + E_2 + \cdots + E_k,\\
A = \lambda_1 E_1 + \lambda_2 E_2 + \cdots + \lambda_k E_k.
\end{gather*}
Adem\'as, el entero $k$, los proyectores ortogonales y los n\'umeros reales
(con las propiedades indicadas) son \'unicos.
$\quad \blacksquare $\\
Se llama \textit{Teorema Espectral} para matrices auto-adjuntas.
Rec\'iprocamente, si tenemos un entero $k \ge 1 $ y proyectores ortogonales y n\'umeros reales
con las propiedades indicadas y luego definimos
$$
A: = \lambda_1 E_1 + \lambda_2 E_2 + \cdots + \lambda_k E_k,
$$
resulta que $A$ es una matriz auto-adjunta.
Esta representaci\'on de $A$ se conoce como su \textit{resoluci\'on espectral}.
Es una \textit{forma can\'onica} de $A$.

Es importante notar que hay una biyecci\'on: $\lambda_j \longleftrightarrow E_j$.

\section{M\'as del Mundo Cl\'asico}

\noindent
Vamos a considerar nuevamente la teor\'ia cl\'asica
de $\Omega$ finito y no vac\'io con $n = \mathrm{card}(\Omega) \ge 1$, pero
ahora sin funci\'on de probabilidad $P$.

Para cada evento $\Lambda \subset \Omega$ definimos su
\textit{funci\'on caracter\'istica} $\chi_\Lambda$ por
$$
       \chi_\Lambda(\omega) := \Big\lbrace
\begin{array}{l}
   1 \mathrm{~si~} \omega \in \Lambda,\\
   0 \mathrm{~si~} \omega \in \Omega, \, \omega \notin \Lambda.
\end{array}
$$

Se sigue que
$\chi_\Lambda = \chi_\Lambda^* = \chi_\Lambda^2 $ y adem\'as $\Lambda=\chi_{\Lambda}^{-1}(1)$.
(Aqu\'i $\chi_\Lambda^2 = \chi_\Lambda \cdot \chi_\Lambda $,
el producto conmutativo usual de funciones como ya hemos definido.)

Adem\'as resulta que cada funci\'on $\chi : \Omega \to \mathbb{C}$ que satisface
$\chi = \chi^* = \chi^2 $
es la funci\'on caracter\'istica de un evento \'unico $\Lambda$,
a saber, $\chi = \chi_\Lambda$ donde $\Lambda = \chi^{-1} (1)$.
En fin de cuentas tenemos una correspondencia uno-a-uno y sobre (o sea, una biyecci\'on)
entre el conjunto de eventos $\Lambda$ en $\Omega$
y el conjunto de las funciones caracter\'isticas con dominio $\Omega$:
$\quad \Lambda \longleftrightarrow \chi_{\Lambda}$.

Adem\'as supongamos que $A: \Omega \to \mathbb{R}$ es una funci\'on arbitraria
(variable aleatoria). Entonces su rango satisface
$
\mathrm{Ran}(A) = \{ \lambda_1, \lambda_2, \dots , \lambda_k  \}
$
\noindent
para un entero $k$ tal que $1 \le k \le n$ y para  $\lambda_1$, $\lambda_2$, $\dots$, $\lambda_k$
  \textit{n\'umeros reales distintos}.

\textbf{N.B.} Si definimos el espectro de la funci\'on $A$ (notaci\'on: $\mathrm{SPEC}(A)$)
como el conjunto de n\'umeros reales
$\lambda$ tal que la funci\'on $\lambda -A $ no tiene inversa multiplicativa,
entonces $\mathrm{SPEC}(A) = \mathrm{Ran}(A)$.

Para cada ``eigenvalor'' $\lambda_j \in \mathrm{SPEC}(A)$ definimos el evento
$$
 \Lambda_j := \mathrm{ker} (\lambda_j - A)=
      \{ \omega \in \Omega ~|~ A(\omega) =\lambda_j \} = A^{-1} (\lambda_j ) \ne \emptyset.
$$
Se dice que $\Lambda_{j}$ es el evento donde (o cuando) $A$ \textit{tiene el valor} $\lambda_j$.
(Se tiene que $A^{-1} (\lambda ) = \emptyset$ si $\lambda \notin \mathrm{SPEC}(A)$.)
Se sigue que son \textit{eventos disjuntos} con
$$
\Omega = \bigcup_{j=1}^k \, \Lambda_j.
$$

Entonces, $E_j := \chi_{\Lambda_j} \ne 0$ para cada $j=1,2, \dots, k$.
Otra vez hay una biyecci\'on importante: $\lambda_j \longleftrightarrow E_j$.
Adem\'as,
\begin{gather*}
 E_i E_j = 0 \quad \mathrm{si} \quad i \ne j, \\
1 = E_1 + E_2 + \cdots + E_k,\\
A = \lambda_1 E_1 + \lambda_2 E_2 + \cdots + \lambda_k E_k.
\end{gather*}
Se llama la \textit{forma can\'onica} de la funci\'on simple $A$.
Es un lema b\'asico en un curso introductorio de la medida.
Hay que recordar que una \textit{funci\'on simple} es una funci\'on (medible)
cuyo rango tiene un n\'umero finito de valores.
Por cierto, este lema no dice nada sobre medidas, como el
teorema espectral para matrices auto-adjuntas no dice nada sobre
medidas de probabilidad cu\'antica.

Entonces hay la misma estructura algebraica en este caso conmutativo
de funciones como en el caso no conmutativo de matrices.
(A veces se dice que \textit{``Cuantizaci\'on es Operadores en lugar de Funciones}''.
No es un teorema tal cual. Resulta que hay muchas cuantizaciones.
Pero es una idea
que puede ser muy \'util, a\'un m\'as \'util que unos teoremas.)

\section{Regresando al Mundo Cu\'antico}

\noindent
Sabiendo como interpretar la forma can\'onica de una funci\'on simple,
podemos dar por analog\'ia
la intepretaci\'on f\'isica en la
 mec\'anica cu\'antica de una matriz auto-adjunta $A = A^*$.
Primero, cada eigenvalor $\lambda_j$ de $A$ es un \textit{resultado posible} de un experimento
que mide la cantidad f\'isica correspondiente a $A$, y no hay
otros resultados posibles.

Tambi\'en el proyector ortogonal asociado $E_j$ debe ser
el evento cu\'antico que corresponde a la medici\'on de $\lambda_j$.
Por lo tanto, definimos
un \textit{Evento (Cu\'antico)} como un proyector ortogonal $E$ en $\mathrm{MAT}(n; \mathbb{C})$,
o m\'as bien, un subespacio $V$ de $\mathbb{C}^n $.

\vskip .2cm \noindent
\textbf{Ejemplo:}
Las matrices $ \, \frac{1}{2}  \sigma_1, \, \frac{1}{2} \sigma_2, \, \frac{1}{2} \sigma_3 $
corresponden a los tres componentes en direcciones ortogonales en el
espacio euclideano $\mathbb{R}^3$ del \textit{spin} de un sistema f\'isico con spin $1/2$.
(Para aprender la \textit{f\'isica} de spin, hay que leer un texto de f\'isica cu\'antica.
Aqu\'i hablamos solamente de la \textit{matem\'atica} de spin.)

\vskip 0.2cm \noindent
\textbf{Ejercicio}: Encontrar la resoluci\'on espectral
para estas tres matrices.

\vskip 0.2cm
Si  $E$ es un evento con $\dim_{\mathbb{C}} \mathrm{Ran} (E) =1 $,
se dice que es un \textit{Estado} o un \textit{Estado Cu\'antico (Puro)}.
Equivalentemente, un estado es un subespacio $V \subset \mathbb{C}^n $ de dimensi\'on uno.
(Es dif\'icil explicar esta definici\'on sin mencionar la evoluci\'on temporal
de un sistema cu\'antico, cosa que no haremos.)

En f\'isica se dice que un estado es un vector $z \in \mathbb{C}^n $ con
$ ||z|| =1 $, porque
este vector define el subespacio $\mathbb{C}  z  \subset \mathbb{C}^n$ de dimensi\'on uno.
Por cierto, hay que \textit{identificar} vectores $w,z \in \mathbb{C}^n$ con $ ||w || = ||z|| =1$
$\Longleftrightarrow$ $\mathbb{C} w = \mathbb{C} z \Longleftrightarrow $
existe  $\alpha \in \mathbb{C}$ con $|\alpha| =1$ tal que $w = \alpha z$.
(Aqu\'i $\mathbb{C}z := \{ \lambda z \, | \, \lambda \in \mathbb{C} \}$.)

Los expertos ya saben que el conjunto de los estados cu\'anticos puros es el
\textit{espacio proyectivo complejo} de
dimensi\'on $n-1$ sobre los complejos: $\mathbb{C} P^{n-1}$.
Para los que todav\'ia no son expertos recomendamos nuevamente la \textit{lectura}.

\vskip 0.1cm
 Ya tenemos el lenguaje suficiente para describir la Probabilidad Cu\'antica.

\vskip 0.1cm \noindent
\textbf{Un Principio de la Mec\'anica Cua\'ntica:}\\
Si un sistema cu\'antico empieza en un estado $z \in \mathbb{C}^n $ con $|| z || =1$
y se mide la cantidad
f\'isica asociada a la matriz auto-adjunta
$$
A = A^* = \lambda_1 E_1 + \lambda_2 E_2 + \cdots + \lambda_k E_k,
$$
escrita en su resoluci\'on espectral,
entonces el experimento da el valor $\lambda_j $ con \textit{Frecuencia Relativa}
$$
\left\langle  z, E_j z \right\rangle = || E_j z ||^2.
$$
\vskip .2cm \noindent
(Despu\'es de la medici\'on de $\lambda_j$, el estado final es
$E_j z / || E_j z  ||$, un cambio que se llama:
\textit{el colapso de la funci\'on de onda}.
Pero no usaremos esto.) $\quad \blacksquare $

\vskip .1cm \noindent
Primero, debemos notar que los n\'umeros $\langle  z, E_j z \rangle $ para $j=1, \dots ,k$ satisfacen
de hecho las propiedades siguientes de frecuencias relativas:
$$
    0 \le \left\langle  z, E_j z \right\rangle \le 1 \quad \quad \mathrm{y} \quad \quad
\sum_{j=1}^k \left\langle  z, E_j z \right\rangle =1.
$$
Si se mide $\lambda_j$ se dice que el evento cu\'antico $E_j $ ha \textit{sucedido}.
Pero cada evento cu\'antico $E$ se encuentra en la resoluci\'on espectral de alguna
matriz auto-adjunta, digamos la matriz $E=E^*$ misma.
Entonces, tenemos que definir la probabilidad para cada evento, no meramente para
los eventos en la resoluci\'on espectral de $A$.
Por lo tanto, definimos la
 \textit{Probabilidad Cu\'antica} para que suceda el evento cu\'antico $E$ (arbitrario)
en el estado $z  \in \mathbb{C}^n$ con $|| z  || =1$ por
$$ \mathrm{Prob} (E; z) := \left\langle  z, E z \right\rangle = || E z ||^2.
$$

\vskip .1cm \noindent
Es important\'isimo notar que el mapeo
$E \to \left\langle  z, E z \right\rangle  \in [0,1]$  (con el estado $z$ fijo)
\textit{no} es una funci\'on de probabilidad cl\'asica (de Kolmogorov) si $n \ge 2$.

\noindent
Supongamos que tenemos una matriz auto-adjunta $A=A^*= \sum_{j=1}^k \lambda_j E_j$
(en la resoluci\'on espectral)
y un estado $z \in \mathbb{C}^n $, o sea, $ || z || =1 $.
Entonces el \textit{Valor Esperado de $A$ en el estado $z$} est\'a definido por
\begin{gather*}
 \left\langle A \right\rangle_z :=
\sum_{j=1}^k \lambda_j \left\langle z, E_j z \right\rangle
= \langle z, \sum_{j=1}^k \lambda_j E_j z \rangle = \left\langle z , A z \right\rangle.
\end{gather*}

Esto nos da otra manera de pensar en
que es un estado $z$.
Se define el mapeo
lineal $\rho : \mathcal{A} \to \mathbb{C}  $
donde $\mathcal{A} =  \mathrm{MAT}(n; \mathbb{C})$ con un estado $z$ fijo por
$$
      \rho(M) := \left\langle z, M z \right\rangle \quad \mathrm{para} \quad M \in \mathcal{A}.
$$
Entonces, tenemos las dos propiedades siguientes.\\
$\rho$ es \textit{positivo:} $\rho(M^*M) \ge 0 \qquad \mathrm{para~toda} \, \, M \in \mathcal{A}$.\\
$\rho$ es \textit{normalizado:} $\quad \rho(I)=1 $.
\vskip .1cm \noindent
As\'i se puede generalizar la idea de un
estado al contexto de una *-\'algebra $\mathcal{A}$ sobre $\mathbb{C}$ con unidad $I$ que
es un \'algebra sobre los complejos $\mathbb{C}$ con un mapeo
$A \mapsto A^* \in \mathcal{A} $ para cada $A \in \mathcal{A}$ tal que para $A,B \in \mathcal{A}\,$ y
$\, \lambda \in \mathbb{C}$ tenemos
\begin{enumerate}
\item $(A+B)^* = A^* + B^*$,
\item $(\lambda A)^* = \lambda^* A^*$,
\item $(AB)^* = B^* A^*$,
\item $A^{**} = A$.
\end{enumerate}
Un \textit{Estado} de una *-\'algebra $\mathcal{A}$ (de dimensi\'on finita, digamos)
sobre $\mathbb{C}$ es un mapeo lineal
$\rho : \mathcal{A} \to \mathbb{C}  $ que es positivo y normalizado (como arriba).
Los elementos $A \in \mathcal{A}$ tal que $A = A^*$ (que se llaman los elementos
\textit{Hermitianos} o \textit{Auto-Adjuntos}) corresponden a las matrices auto-adjuntas
en la mec\'anica cu\'antica y, como veremos al rato, a las variables aleatorias en la probabilidad cl\'asica.
(A veces se dice que cada elemento $A \in \mathcal{A}$ es una variable aleatoria.)

La probabilidad cl\'asica $(\Omega, P)$ es el caso cuando
$\mathcal{A} = \{ Y ~|~ Y : \Omega \to \mathbb{C}  \}$
es la *-\'algebra sobre $\mathbb{C}$
y $\rho (Y) = \sum_{\omega \in \Omega} P(\omega) Y(\omega)$ es el estado.
Entonces los elementos auto-adjuntos en $\mathcal{A}$ son las variables aleatorias.
Adem\'as se suele decir que $P$ es el estado, debido que $\rho$ determina
$P$ \'unicamente.

\section{Conclusi\'on}

\noindent
Entonces la \textit{Probabilidad Cu\'antica} (que empez\'o con los trabajos de Murray y von Neumann
en los 1930's) en nuestro caso de dimensi\'on finita consta de:
\begin{itemize}
 \item El espacio vectorial $\mathbb{C}^n $ con su producto interior.
\item Los eventos cu\'anticos.
\item Los estados cu\'anticos.
\item Las matrices auto-adjuntas.
\item Las f\'ormulas para calcular las probabilidades cu\'anticas, \\
valores esperados, etc\'etera.
\end{itemize}

\section{El Porvenir}

La finalidad de este art\'iculo es darle al lector las ganas para
seguir adelante con estudios de la probabilidad cu\'antica y temas relacionados.
Damos ahora unas referencias, pero es una lista peque\~na.
Son nuestras preferencias. Hay muchas otras referencias buenas.

Antes que nada vale la pena aprender un poco de la f\'isica cu\'antica.
Una introducci\'on muy intuitiva es \cite{FEYN}.

Unas referencias para la probabilidad cu\'antica son \cite{DAV}, \cite{HOL}, \cite{MEY} y \cite{RED}.
La \'ultima referencia muestra claramente que hay muchas probabilidades no cl\'asicas.

Luego hay que estudiar la teor\'ia espectral en dimensi\'on infinita.
Para eso uno puede leer \cite{vN} o volumen I de \cite{RS}.

Para la ecuaci\'on de Schr\"odinger, v\'ease \cite{SUD} para un nivel introductorio
y los vol\'umenes II, III y IV de Reed y Simon \cite{RS} para un nivel avanzado.
Para otros modelos de evoluci\'on temporal, incluyendo los procesos cu\'anticos estoc\'asticos,
v\'ease \cite{CHEB}.

Para una introducci\'on a spin (y bosones y fermiones), v\'ease \cite{SUD}.

Para el c\'alculo cu\'antico estoc\'astico, hay el texto \cite{PAR}.

Para la probabilidad libre, que
es una teor\'ia de probabilidad no cl\'asica muy estudiada,
recomendamos \cite{HP} y \cite{VDN},

Por cierto, hay a\'un m\'as temas interesantes como independencia, \'algebras de von Neumann,
teoremas de l\'imite central, espacios de Fock, movimiento Browniano,
etc\'etera.
Pero el lector debe descubrir el placer de buscar por su propia cuenta
en bibliotecas, librer\'ias e Internet.

\section{Ap\'endice: Qubits}

\vskip .4cm \noindent
Un caso interesante de la mec\'anica cu\'antica es $n=2$,
donde tenemos que el espacio de estados cu\'anticos puros es
$
       \mathbb{C} P^{1} \cong S^{3}/S^{1}  \cong S^{2}.
$
(Aqu\'i $S^n$ es la esfera de dimensi\'on $n$ de vectores de norma uno en $\mathbb{R}^{n+1}$.)

\vskip .1cm \noindent
Un estado tal se llama \textit{Qubit}
en la \textit{Computaci\'on Cu\'antica} y en la \textit{Informaci\'on Cu\'antica}.
Entonces hay una esfera de qubits.

\vskip .1cm \noindent
En el caso $n=2$ de probabilidad cl\'asica, tenemos $\, \Omega = \{ \uparrow, \downarrow \}$,
digamos, que tiene dos estados puros
que son $\{\uparrow \}$ y
$\{ \downarrow \}$; cada uno se llama un \textit{Bit}.
Entonces hay solamente dos bits.

\vskip .1cm \noindent
Por eso hay una diferencia muy grande entre la \textit{computaci\'on cu\'antica} basada en qubits
y la \textit{computaci\'on cl\'asica} basada en bits.
Por ejemplo hay \textit{algoritmos cu\'anticos} much\'isimo
m\'as r\'apidos que los correspondientes \textit{algoritmos cl\'asicos} conocidos.
Hay mucha investigaci\'on en estas \'areas con muchos problemas abiertos.
V\'ease \cite{PETZ} y \cite{RUSK}.

\section{Agradecimientos}
Este art\'iculo empez\'o como una conferencia que impart\'i en el evento
``M\'etodos Estoc\'asticos en Sistemas Din\'amicos, La Probabilidad y su interacci\'on
con otras areas de la Matem\'atica''.
Tuvo lugar
en el Centro de Investigaci\'on en Matem\'aticas (CIMAT),
Guanajuato, Gto., M\'exico,
26-30 de enero de~2009.
Quiero agradecer a los organizadores
(Xavier G\'omez Mont, Renato Iturriaga, Jos\'e Alfredo L\'opez Mimbela,
Joaqu\'in Ortega y Ekaterina Todorova) por su invitaci\'on tan amable a
participar en ese evento.
Por su ayuda con el uso del espa\~nol, le doy muchas gracias a Lenin Echavarr\'ia Cepeda.
Tambi\'en quiero subrayar que el entusiamo, inter\'es y apoyo de Luigi Accardi
son cosas sin las cuales nunca habr\'ia yo empezado a aprender las maravillas de la probabilidad cu\'antica.
Molte grazie, Gigi.

\end{document}